\documentclass{jetpl}

\usepackage{graphicx}
\usepackage{cite}

\twocolumn \lat

\DeclareMathOperator{\sgn}{sgn}

 \title{Decoherence due to nodal quasiparticles in \textit{d}-wave qubits}
 \rtitle{Decoherence due to nodal quasiparticles in \textit{d}-wave qubits}
 \sodtitle{Decoherence due to nodal quasiparticles in \textit{d}-wave qubits}

 \author{Ya.\,V.~Fominov$^{\,+\,*}$\/\thanks{e-mail: fominov@landau.ac.ru, a.golubov@tn.utwente.nl,
                                                     mkupr@pn.sinp.msu.ru},
         A.\,A.~Golubov$^{\,*\,1)}$,
         M.\,Yu.~Kupriyanov$^{\,\square\,1)}$}
 \rauthor{Ya.\,V.~Fominov, A.\,A.~Golubov, and M.\,Yu.~Kupriyanov}
 \sodauthor{Fominov, Golubov, and Kupriyanov}

\address{
$^+$ L.\,D.~Landau Institute for Theoretical Physics RAS, 117940 Moscow, Russia\\
$^*$ Department of Applied Physics, University of Twente, 7500 AE Enschede, The Netherlands\\
$^\square$ Nuclear Physics Institute, Moscow State University, 119992 Moscow, Russia}

\dates{18 May 2003}{*}

\abstract{We study the Josephson junction between two \textit{d}-wave superconductors, which is discussed as an
implementation of a qubit. We propose an approach that allows to calculate the decoherence time due to an intrinsic
dissipative process: quantum tunneling between the two minima of the double-well potential excites nodal quasiparticles
which lead to incoherent damping of quantum oscillations. The decoherence is weakest in the mirror junction, where the
contribution of nodal quasiparticles corresponds to the superohmic dissipation and becomes small at small tunnel
splitting of the energy level in the double-well potential. For available experimental data, we estimate the quality
factor.}

\PACS{85.25.Cp, 85.25.Hv, 74.50.+r, 73.23.-b}

\begin{document}
\maketitle

\begin{figure}
 \centerline{\includegraphics[width=77mm]{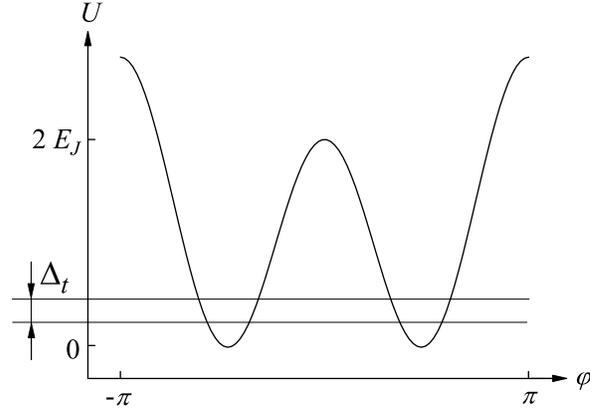}}
\caption{Fig.\ref{fig:2well}. Schematic dependence of the Josephson energy $U$ on the phase difference $\varphi$ (in the
flux qubit $\varphi$ is substituted by $2\pi \Phi/ \Phi_0$, with $\Phi_0$ the flux quantum). The barrier of the height
$2 E_J$ separates two nontrivial minima. The splitting of the lowest energy level due to the tunneling across the
barrier is denoted $\Delta_t$.}
 \label{fig:2well}
\end{figure}

Among various candidates for physical implementation of quantum bits, solid-state proposals, and in particular
superconducting devices, have a number of advantages, e.g., scalability and variability \cite{MSS}. Particularly
interesting are the so-called quiet qubits, which are \textit{intrinsically} degenerate, i.e., do not require any
external source for maintaining the degeneracy. Such qubits can be realized in systems involving \textit{d}-wave
superconductors \cite{Ioffe}. Recently, it was experimentally demonstrated that a double-well potential is indeed
realized in the Josephson junctions between \textit{d}-wave superconductors \cite{Il'ichev}. The qubit variable in this
case is the phase difference $\varphi$ across the junction. The energy of the \textit{phase} qubit has two nontrivial
minima as a function of the phase difference (see Fig.\ref{fig:2well}). Alternatively, a quiet \textit{flux} qubit can
be realized if the spontaneous flux is generated in the loop of \textit{d}-wave superconductors \cite{Tsuei}. The two
qubit implementations are quite similar; for definiteness we shall speak about the phase qubit.

In such intrinsic qubits, there are also intrinsic mechanisms of decoherence even at low temperatures. The quantum
tunneling of the phase between the two minima leads to fluctuating voltage across the junction, which excites
quasiparticles. The dissipative current across the interface arises, leading to a finite decoherence time
$\tau_\varphi$. The knowledge of $\tau_\varphi$ is essential for estimating the efficiency of the qubit: short
decoherence time makes the qubit senseless, while a long enough decoherence time opens the way for quantum correction
algorithms that in principle allow to perform an infinitely long computation \cite{corr}.

The relevance of the quasiparticle processes at low temperatures is specific for \textit{d}-wave superconductors. In the
conventional \textit{s}-wave case, the quasiparticle transport below the gap is suppressed. At the same time, in gapless
anisotropic superconductors the gap vanishes in certain directions (the nodal directions), hence the low-energy
quasiparticle appear. In the present letter, we consider a DID Josephson junction (D = \textit{d}-wave superconductor, I
= insulator), and study the decoherence due to nodal quasiparticles (quasiparticles moving along the nodal directions).

\textbf{Decoherence time (general strategy).} Theoretical description of the quantum dynamics of a tunnel junction
between two \textit{s}-wave superconductors was developed in Ref. \cite{ESA} (see Ref. \cite{SZ} for a review). The
effective action for the phase difference $\varphi$ was obtained. Later this description was generalized to the case of
\textit{d}-wave superconductors in Refs. \cite{Bruder,BGZ}. The effective action for $\varphi$ is similar to the general
case considered by Caldeira and Leggett \cite{CL,Leggett_review}, who studied influence of dissipation on quantum
tunneling in macroscopic systems. The dissipation was described as being due to the interaction with a bath of
oscillators (the environment). The ``strength'' of the environment, depending on the frequency $\omega$, is
characterized by the spectral function $J(\omega)$. In the Josephson junction, the environment is represented by the
quasiparticles, and the spectral function is given by $\hbar I(\hbar\omega/e) /e$, where $I$ is the dissipative
quasiparticle current taken at ``voltage'' $\hbar \omega /e$ \cite{ESA}.

A system living in a double-well potential and described by an extended coordinate can be ``truncated'' to the two-state
system (spin $1/2$) with the two states ($\sigma_z =\pm 1$) corresponding to the minima of the potential (see
Fig.\ref{fig:2well}). The theory of dissipative two-state systems is thoroughly elaborated \cite{Leggett_review} for the
cases when the spectral function behaves as $J(\omega) \propto \omega^s$ up to some high-frequency cutoff. The
situations when $s=1$, $s>1$, and $0<s<1$ are called ohmic, superohmic, and subohmic, respectively. In this language,
the dissipation due to nodal quasiparticles in the Josephson junction is superohmic, as we demonstrate below.

What is the decoherence in such a system? Assume that during the time $t<0$ the system is held in the right well (i.e.,
at $\sigma_z=1$). At $t=0$ the constraint is released, and we consider the expectation value of the system coordinate:
$P(t) = \left< \sigma_z(t) \right>$. Below we shall encounter the superohmic case at zero temperature. Then
\cite{Leggett_review}
\begin{equation}
P(t) = \cos(\Delta_t t/\hbar) \exp(- t / \tau_\varphi)
\end{equation}
--- the cosine describes coherent oscillations between the two wells ($\Delta_t$ is the tunnel splitting of levels,
see Fig.\ref{fig:2well}) while the exponential leads to their incoherent damping.

The decoherence time $\tau_\varphi$ is expressed in terms of the spectral function \cite{Leggett_review}. Returning from
the general theory to the particular case of the Josephson junction, we write the corresponding result as
\begin{equation}
\tau_\varphi = \frac{4e}{\delta\varphi^2 I(\Delta_t/e)} =\frac{4\pi\hbar}{\delta\varphi^2 e R_q I(\Delta_t/e)},
\end{equation}
where $\delta\varphi$ is the distance between the potential minima and $R_q = h/2 e^2 \approx 13\,\mathrm{k}\Omega$ is
the quantum resistance. Comparing the decoherence time with the characteristic time of oscillations between the wells,
$\hbar/ \Delta_t$, we obtain the quality factor
\begin{equation} \label{Q}
Q = \frac{\tau_\varphi \Delta_t}{2\hbar} = \frac{2\pi \Delta_t} {\delta\varphi^2 e R_q I(\Delta_t/e)},
\end{equation}
which must be large for successful operating of the qubit.

In the DID junction, the tunnel splitting $\Delta_t$ is much smaller than the order parameter $\Delta$, hence
$\tau_\varphi$ is determined by the quasiparticle current at low ``voltage''.

\begin{figure}
 \centerline{\includegraphics[width=84mm]{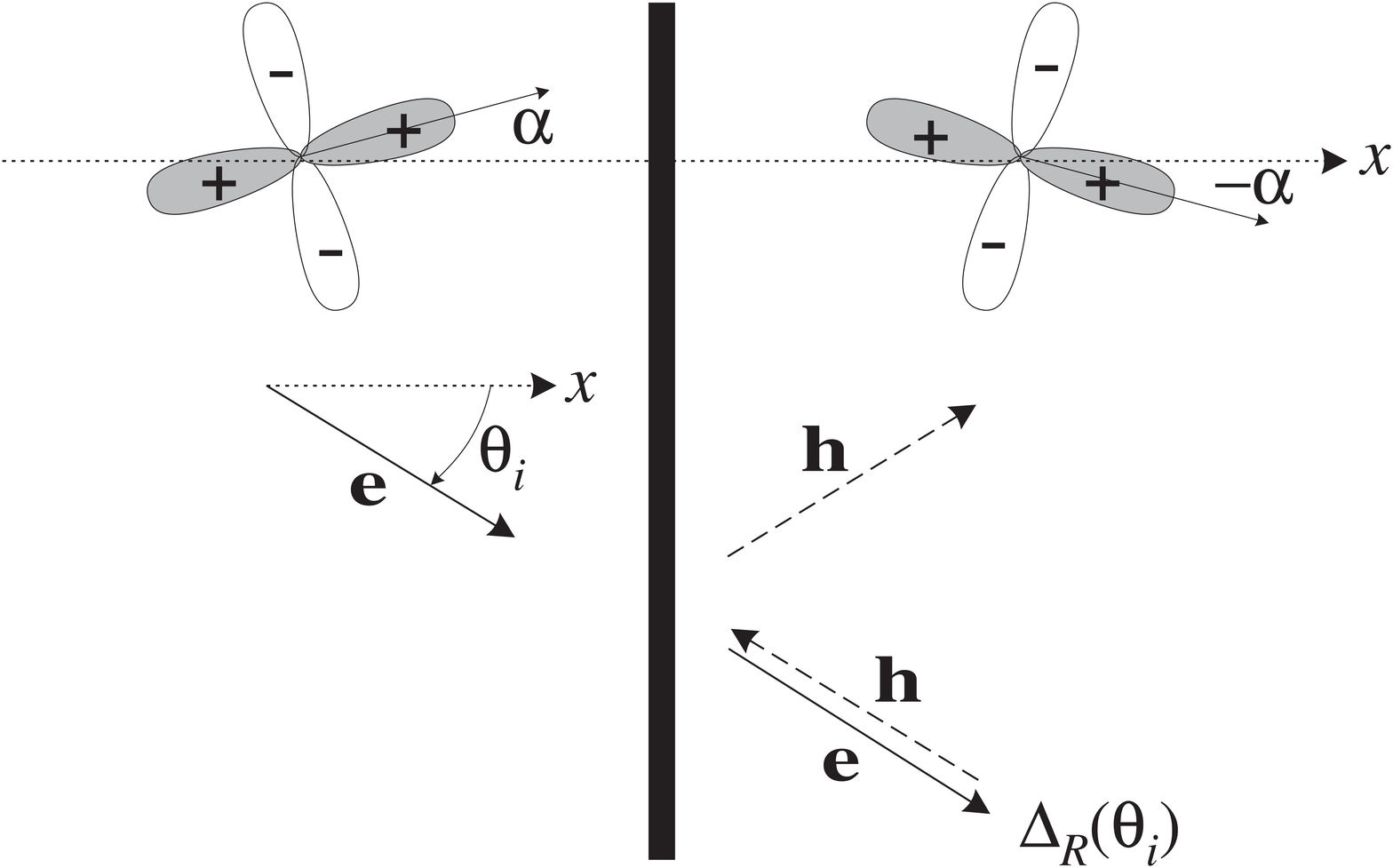}}
\caption{Fig.\ref{fig:system}. DID junction of mirror orientation $\alpha/-\alpha$.
An electron \textbf{e} moving along a truly nodal direction $\theta_i$ of the left superconductor,
tunnels into an induced nodal direction of the right superconductor. $\Delta_R (\theta_i) \ne 0$, therefore the electron
experiences the Andreev reflection; the hole \textbf{h} returns to the interface, and after reflection at the interface
escapes into the bulk along the truly nodal direction $-\theta_i$. In this process, the total current into the bulk of
the right superconductor is composed of the Cooper pair along $\theta_i$ and the hole along $-\theta_i$.}
 \label{fig:system}
\end{figure}

\textbf{Quasiparticle current.} Motivated by the experiment \cite{Il'ichev}, we consider the grain-boundary Josephson
junction between two quasi-two-dimensional $d_{x^2-y^2}$-wave superconductors with cylindrical Fermi surfaces. The
orientations of the superconductors are characterized by the angles between the \textit{a}-axes and the normal to the
interface (the $x$-axis) --- see Fig.\ref{fig:system}. According to Ref. \cite{Il'ichev}, we consider the mirror
junction, in which the misorientation angles on both sides are equal in magnitude but opposite in sign, $\alpha/-\alpha$
(we take $-45^\circ \leqslant \alpha \leqslant 45^\circ$ because all physically different situations in the mirror
junction are realized in this interval). The order parameter depends on the direction (parametrized by the angle
$\theta$) and the distance to the interface:
\begin{equation}
\Delta_{L,R} (x,\theta) = \widetilde\Delta_{L,R} (x) e^{i\varphi_{L,R}} \cos\left( 2 (\theta \mp \alpha) \right),
\end{equation}
where the indices $L$ and $R$ refer to the left- and right-hand side of the junction, respectively.

The quasiparticle current in the tunneling limit at low temperatures, $k_B T\ll \hbar\omega$, is given by
\begin{gather}
I(\hbar\omega/e) = \frac 1{e R_N} \int_{-\pi /2}^{\pi /2} d\theta \frac{D(\theta) \cos\theta}{\widetilde
D} \int_0^{\hbar\omega} dE \times \notag \\
\times N \left( E- \hbar\omega,\theta \right) N \left(E,\theta \right), \label{I_V_general}\\
\widetilde D = \int_{-\pi/2}^{\pi/2} d\theta D(\theta)\cos\theta. \notag
\end{gather}
Here $R_N$ is the normal-state resistance of the interface, $N(E,\theta)$ is the density of states (DoS) at the
interface, normalized to the normal-metal value, and $D(\theta)$ is the angle-dependent transparency of the interface.
We have not labelled the DoS by the indices $L$ and $R$ because $N_L(E,\theta) =N_R(E,\theta)$ in the mirror junction.

Below we calculate the nodal contribution to the current (\ref{I_V_general}) at $\hbar\omega \ll \widetilde\Delta_0$,
where $\widetilde\Delta_0 = \widetilde\Delta(\pm\infty)$ is the bulk amplitude of the order parameter. The angle
integration contributing to the current is then limited to narrow angles around the nodal directions, where the
low-energy DoS is nonzero (as we shall see below, the width of the angles is $\delta\theta = \hbar\omega
/\widetilde\Delta_0$).

To calculate the DoS, we employ the quasiclassical approach. The quasiclassical matrix Green function
\begin{equation}
\widehat G = \begin{pmatrix} g & f \\ \Bar f & -g \end{pmatrix}
\end{equation}
obeys the Eilenberger equation \cite{Eilenberger} and satisfies the normalization condition $\widehat G^2=\widehat 1$.
It can be parametrized as
\begin{equation} \label{parametrization}
g=\frac{1-ab}{1+ab},\qquad f=\frac{2a}{1+ab},\qquad \Bar f =\frac{2b}{1+ab},
\end{equation}
then the normalization condition is automatically satisfied. The equations for the new functions $a(x,\theta)$ and
$b(x,\theta)$ take the form of the Riccati equations \cite{Schopohl}:
\begin{gather}
\hbar v_F \cos\theta (da / dx) - 2 i E a +\Delta^* a^2 -\Delta =0, \notag \\
\hbar v_F \cos\theta (db / dx) + 2 i E b -\Delta b^2 +\Delta^* =0, \label{b}
\end{gather}
where $v_F$ is the absolute value of the Fermi velocity $\mathbf{v}_F$, and $\theta$ denotes the angle between
$\mathbf{v}_F$ and the $x$-axis.

In the tunneling limit, the DoS is calculated at an impenetrable interface. Let us consider, e.g., the right
superconductor (the right half-space). We need to find the low-energy DoS in two cases: 1)~in the vicinity of a nodal
direction, so that $E, \Delta(\theta) \ll \widetilde\Delta_0$, 2)~at a gapped direction, so that $E \ll \Delta(\theta)$.
In the first case, the space scale $\xi_E = \hbar v_F \cos\theta / |\mathcal{E}_+|$ on which the quasiclassical Green
functions vary (we denote $\mathcal{E}_\pm = \sqrt{E^2- \left| \Delta (\infty,\pm \theta) \right|^2}$), is much larger
than the coherence length $\xi = \hbar v_F / 2\pi k_B T_c$ on which variations of $\Delta$ occur. This allows us to
regard $\Delta$ as constant when integrating Eqs. (\ref{b}) over $x$. In other words, the functions $a$ and $b$ at low
energies do not feel the suppression of $\Delta$ near the interface, because it takes place on a small scale. In the
second case, the spatially dependent parts of $a$ and $b$ are proportional to $E/\Delta(\theta) \ll 1$ and hence small.
Thus $a$ and $b$ at the interface are equal to their bulk values, as if $\Delta$ was constant.

Thus we can regard $\Delta(x,\theta)$ as equal to the bulk value $\Delta_0(\theta)=\Delta(\infty,\theta)$. The
integration of the functions $a$ and $b$ over $x$ in Eqs. (\ref{b}) is stable only in the directions determined by the
sign of $\cos\theta$. At $\cos\theta >0$, the function $b(x,\theta)$ is stably integrated from $x=\infty$ to the
interface ($x=0$), hence
\begin{equation} \label{b_surface}
b(0,\theta) = b(\infty,\theta) = i ( E-\mathcal{E}_+ \sgn E ) / \Delta_0(\theta).
\end{equation}
At the same time at $\cos\theta >0$, the function $a$ is stably integrated from the interface to $x=\infty$. Therefore
to find $a(0,\theta)$, we consider the trajectory directed along $\pi-\theta$. Since $\cos(\pi-\theta) <0$, the function
$a$ is stably integrated from $x=\infty$ to the interface. Finally, the direction $\pi-\theta$ is converted to $\theta$
upon reflection at the specular interface:
\begin{gather}
a(0,\theta) = a(0,\pi-\theta) =a(\infty,\pi-\theta) = \notag \\
= i ( E-\mathcal{E}_- \sgn E ) / \Delta_0^*(-\theta).
\end{gather}
As a result, the DoS $N = \Real g$ at the interface is
\begin{equation} \label{g_interface}
N (E,\theta) = \Real \frac{|E| \left( \mathcal{E}_+ +\mathcal{E}_- \right)} {E^2 - \Delta_0(\theta) \Delta_0^*(-\theta)
+ \mathcal{E}_+ \mathcal{E}_-}.
\end{equation}
The gap in the spectrum is $E_g(\theta)=\min\left( \left| \Delta_0(\theta) \right|, \left| \Delta_0(-\theta) \right|
\right)$.

The DoS is symmetric, $N(\theta) =N(-\theta)$, because the Green functions are continuous upon reflection. Thus in each
superconductor there are two ``truly'' nodal directions $\theta_i$ ($i=1,2$) in the interval $-\pi/2 <\theta < \pi/2$,
and also two ``induced'' nodal directions $-\theta_i$. Near a nodal direction $E_g(\theta)= 2 \widetilde\Delta_0
|\theta-\theta_i|$. Along a truly nodal direction, the gap vanishes and the DoS is the same as in the normal metal,
$N(E)=1$. For an ``induced'' nodal direction this is so only near the interface.

In the left superconductor, the truly nodal directions are $\theta_{1,2} = \alpha \pm 45^\circ$. Due to the mirror
symmetry, the truly nodal directions of the right superconductor coincide with the induced nodal directions of the left
one, and vice versa. In total, there are four nodal directions in the junction, which are symmetric with respect to the
interface normal.

In this situation, the transport is due to the processes of the following type. An electron moving along a truly nodal
direction $\theta_i$ of the left superconductor, tunnels into an induced nodal direction of the right superconductor
(see Fig.\ref{fig:system}). However, the electron cannot escape into the bulk of the right superconductor because
$\Delta_R (\theta_i) \ne 0$. Therefore the electron experiences the Andreev reflection; the hole returns to the
interface, and after reflection at the interface escapes into the bulk along the truly nodal direction $-\theta_i$. In
this process, the total current into the bulk of the right superconductor is composed of the Cooper pair along
$\theta_i$ and the hole along $-\theta_i$, which is overall equivalent to the transfer of one electron.

The nodal contribution to the current (\ref{I_V_general}) appears only due to integrating in the vicinity of the nodal
directions where $E_g < \hbar \omega$. The DoS near the nodal directions at small energies can be found from Eq.
(\ref{g_interface}). Below we distinguish the general case when $\Delta_0 (\theta) \ne \pm \Delta_0 (-\theta)$, and two
special cases: $\Delta_0(\theta) = \Delta_0(-\theta)$ (at $\alpha = 0^\circ$) and $\Delta_0 (\theta) =-\Delta_0
(-\theta)$ (at $\alpha = 45^\circ$).

At $\alpha=0^\circ$, the truly nodal and induced nodal directions coincide in each superconductor, and Eq.
(\ref{g_interface}) yields the BCS-like DoS:
\begin{equation} \label{nu_0}
N_{0^\circ} (E,\theta) = \Real \Bigl( |E| \Bigl/ \sqrt{E^2- \left| \Delta_0(\theta) \right|^2} \Bigr) .
\end{equation}

At $\alpha=45^\circ$, the truly nodal and induced nodal directions again coincide, and Eq. (\ref{g_interface}) yields
the DoS of the inverse BCS type:
\begin{equation} \label{nu_45}
N_{45^\circ} (E,\theta) = \Real \Bigl( \sqrt{E^2- \left| \Delta_0(\theta) \right|^2} \Bigr/ |E| \Bigr) .
\end{equation}

Finally, if $|\alpha| \gg \hbar\omega / \widetilde\Delta_0$ and $45^\circ - |\alpha| \gg \hbar\omega /
\widetilde\Delta_0$ (i.e., $\alpha$ is not too close to $0^\circ$ and $\pm 45^\circ$), then $\Delta_0(\theta)$ in the
essential angle of the width $\delta\theta =\hbar \omega / \widetilde\Delta_0$ around a nodal direction is much smaller
than $\Delta_0(-\theta)$. Then in the region of energies that contribute to the quasiparticle current, $\left|
\Delta_0(\theta) \right| < |E| < \hbar\omega \ll \left| \Delta_0(-\theta) \right|$, the DoS is again given by the
inverse BCS formula:
\begin{equation} \label{nu_i}
N_g (E, \theta\approx \theta_i) =\Real \Bigl( \sqrt{E^2-\left| \Delta_0(\theta) \right|^2} \Bigr/ |E| \Bigr) .
\end{equation}

\begin{figure}
 \centerline{\includegraphics[width=84mm]{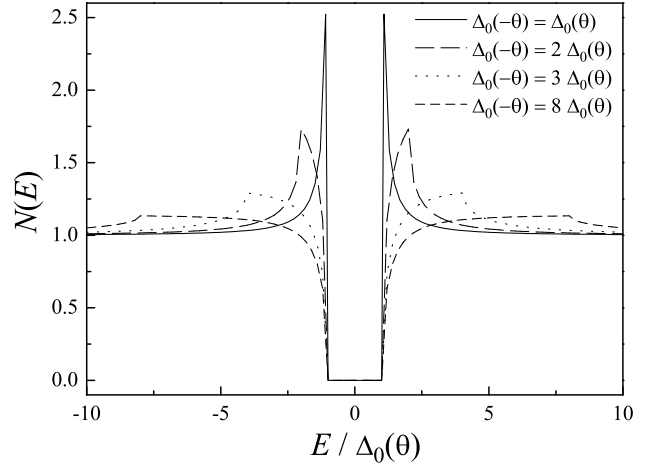}}
\caption{Fig.\ref{fig:dos}. Density of states following from Eq. (\ref{g_interface}). The energy is normalized to
$\Delta_0(\theta)$, while $\Delta_0(-\theta)$ is varied.}
 \label{fig:dos}
\end{figure}

Figure~\ref{fig:dos} demonstrates the DoS at different angles $\theta$, which are parametrized by different ratios
$\Delta_0(-\theta) / \Delta_0(\theta)$. At $\Delta_0(-\theta) = \Delta_0(\theta)$, the DoS has the BCS-like square-root
singularity near $E_g$ [see Eq. (\ref{nu_0})]. At $\Delta_0(-\theta) \ne \Delta_0(\theta)$, the DoS has the inverse-BCS
behavior near $E_g$ [see Eq. (\ref{nu_i})].

Inserting Eqs. (\ref{nu_0})--(\ref{nu_i}) into Eq. (\ref{I_V_general}), we obtain:
\begin{equation} \label{I_V_J}
I(\hbar\omega/e) = \frac{A(\alpha)}{e R_N} \frac{(\hbar\omega)^2}{\widetilde\Delta_0} \sum_{i=1,2} \frac{D(\theta_i)
\cos\theta_i}{\widetilde D},
\end{equation}
where $\theta_{1,2} =\alpha \pm 45^\circ$ and $A$ is a number, which depends on the orientation of crystals: $A(0^\circ)
\approx 0.46$, $A(45^\circ) \approx 0.19$, and $A(\alpha) = 2 A(45^\circ) \approx 0.37$ when $\alpha$ is not too close
to $0^\circ$ or $\pm 45^\circ$.

In Refs. \cite{Bruder,BGZ}, the quadratic current--voltage characteristic, $I\propto \omega^2$, was obtained for the
case of aligned nodal directions (i.e., for the $\alpha/\alpha$ orientation).

\textbf{Estimate.} Equations (\ref{Q}), (\ref{I_V_J}) yield:
\begin{equation} \label{Q_1}
Q = \Biggl( \frac{2\pi} {\delta\varphi^2 A(\alpha) \sum\limits_{i=1,2} D(\theta_i) \cos\theta_i / \widetilde D} \Biggr)
\frac{R_N}{R_q} \frac{\widetilde\Delta_0}{\Delta_t}.
\end{equation}

To proceed further, we need to estimate the tunnel splitting $\Delta_t$ (see Fig.\ref{fig:2well}). For the estimate, we
assume that the second harmonic dominates in the energy--phase relation, $U(\varphi)= E_J (1+\cos 2\varphi)$, and the
energy of the levels is small compared to $E_J$. Then the tunneling action is calculated between the points
$\varphi=-\pi/2$ and $\pi/2$, and we obtain:
\begin{equation} \label{Delta_t}
\Delta_t = ( 4 \sqrt{2 E_J E_C} / \pi) \exp\left( -\sqrt{2 E_J / E_C} \right) ,
\end{equation}
where $E_C = e^2 /2 C$ is the charging energy ($C$ is the capacitance of the junction).

To obtain a numerical estimate, we take the characteristics of the junction as in the experiment of Il'ichev et al.
\cite{Il'ichev}. The capacitance of the junction is $C \sim 10^{-14}$\,F \cite{Grajcar}, hence $E_C / k_B \sim 0.1$\,K.
The characteristic Josephson energy is of the order of several Kelvin. For estimate, we take $2 E_J / k_B = 7$\,K. The
resistance of the interface is $R_N \sim 50\,\Omega$ \cite{Grajcar}.

As a result, $\Delta_t / k_B \sim 2.5\cdot 10^{-4}$\,K. Finally, we estimate $\delta\varphi\sim\pi$, $\widetilde\Delta_0
/ k_B \sim 200$\,K, and assume a thin $\delta$-functional barrier with transparency $D(\theta) = D_0 \cos^2 \theta$,
then the quality factor is $Q \sim 10^3\div 10^4$. Here we have retained only the order of magnitude for $Q$, because we
cannot expect a higher accuracy in the case when important characteristics of the junction (e.g., $C$ and $E_J$) are
known only by the order of magnitude. We also made an essential assumption that the second Josephson harmonic dominates.

The latter assumption can be realized under special conditions, while in a more common situation the first and the
second harmonics are of the same order. Estimates for this case were done in a recent work \cite{Tzalenchuk}, where the
characteristics of mesoscopic junctions between high-$T_c$ superconductors were experimentally studied and theoretically
analyzed. A characteristic value of $\Delta_t \sim 0.1$\,K was reported under the conditions that correspond to $R_N
\sim 100\,\Omega$. Assuming such parameters for the mirror junction, we obtain $Q \sim 10 \div 10^2$.

The above estimates for $Q$ are very different. At the same time, a general consequence of Eq. (\ref{Q_1}) is that the
quality factor grows as the splitting $\Delta_t$ becomes smaller. We note in this respect, that the values of the
critical current (and hence the Josephson energy) measured in Refs. \cite{Il'ichev,Tzalenchuk}, are much smaller than
expected. If the critical current is enhanced to the expected value, then $\Delta_t$ decreases, which finally leads to
an increase of $Q$.

If $\alpha\ne 0^\circ$, the low-energy quasiparticles are presented not only by the nodal quasiparticles, but also by
the midgap states (MGS) with zero energy \cite{Hu}. In the case of specular interface and clean superconductors,
considered in this paper, the DoS corresponding to the MGS is proportional to $\delta(E)$, hence the MGS on the two
sides of the interface do not overlap and do not contribute to the current at a finite voltage.

In the asymmetric case, when $\alpha_L \ne \pm \alpha_R$ (precisely speaking, when $\bigl| |\alpha_L| - |\alpha_R|
\bigr| > \hbar \omega / \widetilde\Delta_0$), the nodal directions of the left and right superconductors do not match
each other. Then the transport from nodal to nodal direction is suppressed. However, a more important transport
``channel'' arises: between the nodal directions and the MGS. This leads to a stronger decoherence than in the symmetric
case.

In the mirror junction, the MGS contribute to the quasiparticle current if they are split and/or broadened
\cite{splitting}. To take into account the contribution of the MGS into decoherence, the present approach should be
considerably modified. This issue requires a separate study.

In conclusion, we have proposed an approach that allows to calculate the decoherence time due to nodal quasiparticles in
the DID junctions, which can be used as phase or flux qubits. The dissipation in the mirror junctions is weaker than in
the asymmetric ones. We find the superohmic dissipation with $s=2$ in the mirror junction, which becomes weak at small
tunnel splitting of the energy level in the double-well potential. For available experimental data, we estimate the
quality factor. The superohmic case is most favorable (compared to ohmic and subohmic) for possible qubit applications.

We thank M.\,V.~Feigel'man for careful reading and comments on the manuscript. We are indebted to A.\,Ya.~Tzalenchuk for
communicating the results of his group to us before publication. We are grateful to M.\,H.\,S.~Amin, Y.~Makhlin,
A.~Shnirman, and A.\,M.~Zagoskin for helpful discussions. The research was supported by the D-Wave Systems Inc. and the
ESF PiShift program. Ya.V.F. was also supported by the RFBR 01-02-17759, the Swiss National Foundation, the Russian
Ministry of Industry, Science and Technology (RMIST), and the program ``Quantum Macrophysics'' of the Russian Academy of
Sciences. M.Yu.K. was also supported by the RMIST.

\end{document}